\documentclass[prl,twocolumn,aps]{revtex4}

\usepackage{graphicx}  % needed for figures
\usepackage{dcolumn}   % needed for some tables
\usepackage{bm}        % for math
\usepackage{amsmath}
\usepackage{placeins}
\usepackage{amssymb}   % for math
\usepackage{setspace} %define space
\usepackage{hyperref}
\hypersetup{
  colorlinks   = true,    % Colours links instead of ugly boxes
  urlcolor     = blue,    % Colour for external hyperlinks
  linkcolor    = blue,    % Colour of internal links
  citecolor    = red      % Colour of citations
}

\begin{document}

\title{Probing light forces on cold atoms by noise correlation spectroscopy}
%in electromagnetically induced transparency}
\author{K. Theophilo, A. Kumar, H. M. Florez, C. Gonz\'alez-Arciniegas, P. Nussenzveig, and M. Martinelli{*}}
\affiliation{Instituto de F\'{\i}sica, Universidade de S\~ao Paulo, 05315-970 S\~ao Paulo, SP-Brazil} 

\begin{abstract}
Enhanced sensitivity in electromagnetically induced transparency (EIT) can be obtained by the use of noise correlation spectroscopy between the fields involved in the process. Here, we investigate EIT in a cold ($< 1$ mK) rubidium vapor and demonstrate sensitivity to detect weak light-induced forces on the atoms. A theoretical model is developed and shows good agreement with our measurements, enabling the attribution of the observed effects to the coupling of the atomic states to their motion. The effects remain unnoticed on the measurement of the mean fields but are clearly manifest in their correlations.

%The sensitivity of spectroscopy in electromagnetically induced transparency can be enhanced by the use of noise correlation spectroscopy between the fields involved. The measurement can become sensitive enough to observe the role of the weak push exerted by the laser fields close to the two photon resonance, probing the coupling of internal and external degrees of freedom in the atomic system. We observe the effect of these processes in the EIT correlation spectroscopy and develop a model for this coupling in good agreement with experimental observations.

\end{abstract}

\email{mmartine@if.usp.br}

\maketitle

Electromagnetically induced transparency (EIT)  \cite{Harris91} is a coherent process that has received  growing  attention in the last decades not only for its applications to quantum networks based on quantum memories \cite{Marangos05,Hugues11}, but also in metrology.
Recent proposals have reported its use for precision measurements in atomic clocks at room temperature, using the EIT narrow linewidth to provide sensitive detection of frequency fluctuations \cite{Novikova17}.  It has also been shown that EIT is well suited for  magnetometry applications \cite{Novikova11} in particular, showing a vector magnetometer operation with angular sensitivity up to $10^{-3}$deg$/\sqrt{Hz}$ \cite{Yudin10}. Moreover,  the use of EIT in diamonds \cite{Acosta13} and Rydberg \cite{Anderson} atoms at room temperature for electrometry applications was recently demonstrated.

In spite of these metrological applications, the sensitivity of  EIT to the external degrees of freedom of the atoms due to light forces has remained somewhat unexplored. 
Nevertheless, the subtle radiation pressure exerted by the pump and probe beams close to resonance modifies the atomic dynamics.
Although this effect is negligible in the spectroscopy of hot atomic vapors, it becomes relevant for measurements in cold gases.
In this paper we show that
we can rely on the sensitivity of the noise correlation spectroscopy \cite{Scully05, Cruz07,  Xiao09} to observe the role of the radiation pressure on the atomic dynamics, otherwise
unnoticed in the average intensity signals.

\begin{figure}[h!]
\centering
\includegraphics[width=50mm]{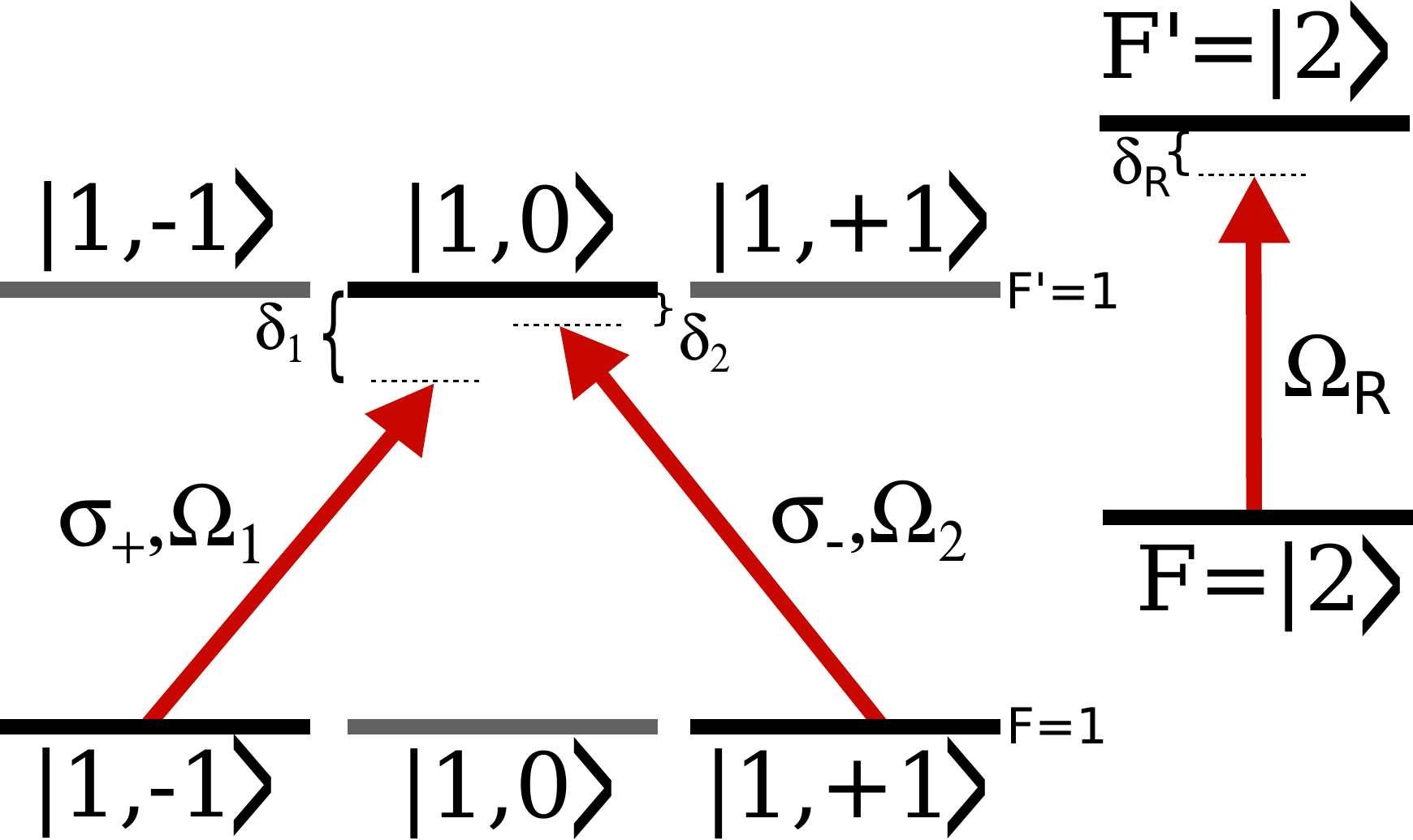}
\caption{ Relevant levels of the $^{87}$Rb atoms under  study (not in scale).}
\label{lambda}
\end{figure}

EIT in a $\Lambda$ transition involves two lower energy states of an atom that are coupled by a pair of fields to an excited state (Fig. \ref{lambda}). As a consequence of this combined interaction, the atoms are pumped into a dark state, a superposition of the ground states that decouples the atoms from the incident fields, therefore leading to a reduction of the light scattering \cite{Harris91}. The transmission of each field presents an increase in the intensity when the $\Lambda$ resonance is satisfied, i. e., the energy diference of the incident photons matches the energy diference of the two levels with lower energy. Atomic dynamics, as well as field fluctuations, disturb the  transmitted intensities, which develop  fluctuations $\delta I_i(t)$.
The analysis of these fluctuations provides a useful tool for the study of atomic dynamics \cite{Felinto13}, but the measured intensity can be affected by other noise sources. 
Their effect can be filtered out by the observation of  the Fourier component of the photocurrent  $\delta I_i(\omega_a)$, and the sensitivity is increased by the use of the  correlation function
\begin{equation}
C(\omega_a) = \frac{S_{12} (\omega_a)}{\sqrt{S_{11}(\omega_a)S_{22}(\omega_a)} }, \,\mbox{where}
\label{eq:Ceq}
\end{equation}
\begin{equation}
S_{ij} (\omega_a)= \frac{1}{2}[\langle\delta I_i(\omega_a)\delta I_j^*(\omega_a)\rangle +\delta I_i^*(\omega_a)\delta I_j(\omega_a)\rangle],
\label{eq:Noiseeq}
\end{equation}
in which $S_{ii} (\omega_a)$ is the noise power of each transmitted beam and $S_{ij} (\omega_a)$ is the cross-correlation of the noise, at the analysis frequency $\omega_a$ \cite{Cruz07}.
This normalization of the correlation makes the result more robust against atomic losses in the system, in contrast to the typical measurement of the intensities of the transmitted beam or the measurement of fluorescence.
 
We studied  an ensemble of cold $^{87}$Rb atoms, investigating the coupling between Zeeman levels in the $5S_{1/2}(F=1)\rightarrow  5P_{3/2}(F'=1)$ transition (energy gap $E=\hbar \omega$), selectively addressed by cross circularly polarized  laser fields with controllable detunings (Fig.\ref{lambda}). Field 1 (Rabi frequency $\Omega_1$) is kept on resonance, while the frequency of the probe field (Rabi frequency $\Omega_2$) is scanned. The one-photon detuning of each field is given by  $\delta_i=(E_i-E)/\hbar$, where $E_i=\hbar\omega_i$ is the photon energy of the laser. Two-photon detuning for the $\Lambda$ transition is defined as $\delta=\delta_2-\delta_1$.
To compensate for the loss of atoms by spontaneous emission into the $F=2$ level, a resonant repump field couples the transition  $5S_{1/2}(F=2)\rightarrow  5P_{3/2}(F'=2)$. The Zeeman levels  with magnetic quantum number $m=\pm1$ in level  $5P_{3/2}(F'=1)$ are not optically coupled by the incident fields, and the transition between the fields with $m=0$ is forbidden, leading therefore to an effectively pure $\Lambda$-system, with a certain rate for the loss and reload of atoms that affects the coherence of the two-photon transition.

%\section{Experimental setup}

\begin{figure}[h!]
\centering
\includegraphics[width=85mm]{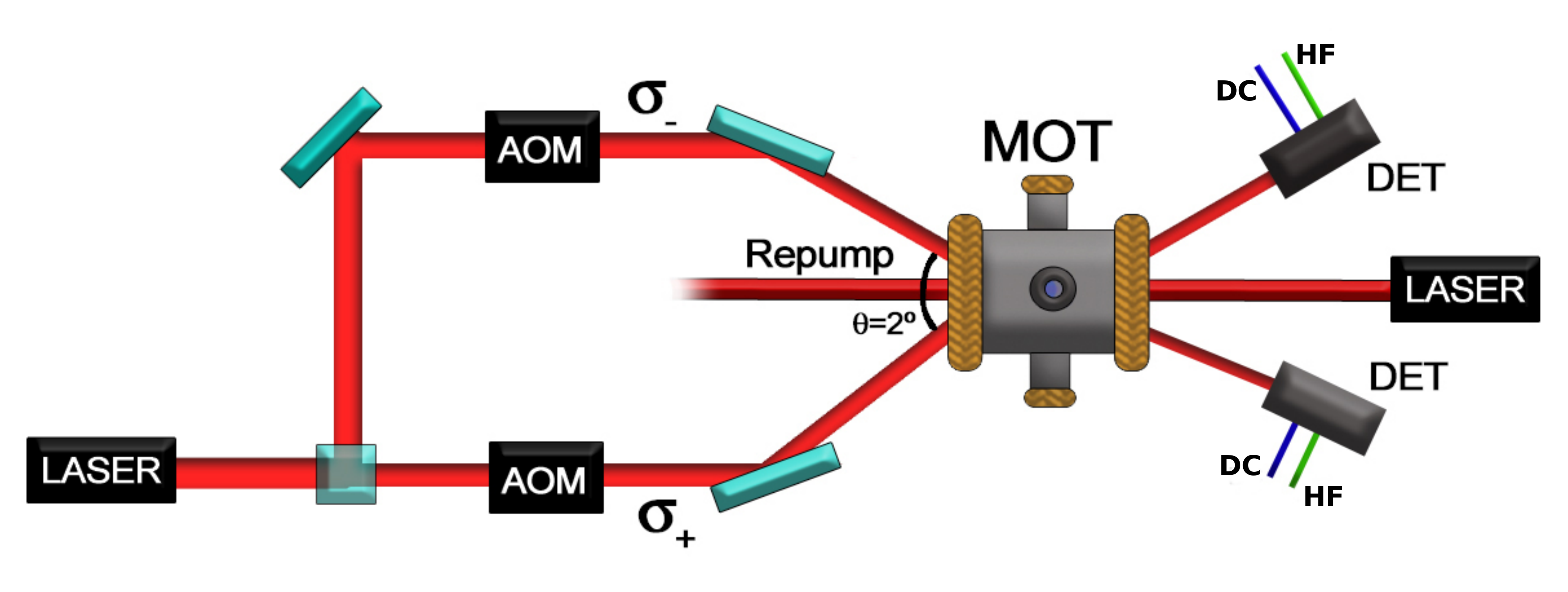}
\includegraphics[width=80mm]{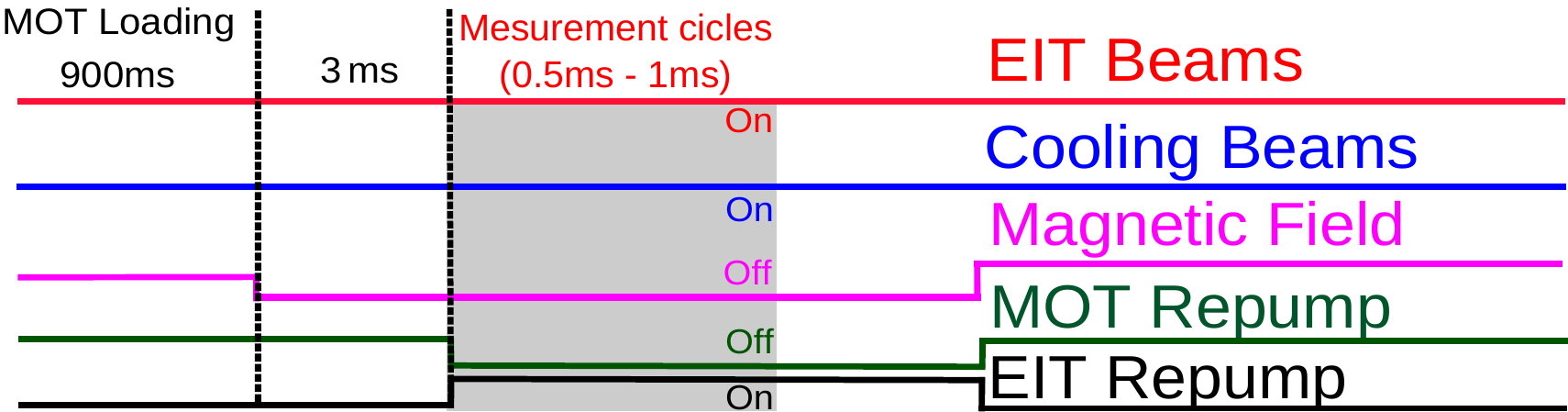}
\caption{ (Top)  Experimental Setup.  (Bottom) Timing sequence of our experimental cycle. }
\label{fig:setup}
\end{figure}

The experimental setup is presented in Fig.\ref{fig:setup}(Top) and it has been previously described in Ref.\cite{Florez16_p1}. It  
consists of an ensemble of cold $^{87}$Rb atoms in a magneto-optical trap (MOT) with nearly $10^7$ atoms at 800 $\mu$K.
 The two cross circularly polarized  beams addressing the EIT transition are nearly
co-propagating (with an angle $\theta$ $\sim$ 2$^{o}$), and their frequencies and powers can be controlled by acousto-optic modulators (AOM).
 The repump laser is anti-parallel with respect to the spectroscopy beams. 
To perform the correlation spectroscopy, we  keep the detuning of the field 1  fixed  
while scanning the two-photon frequency detuning $\delta'=\delta/2\pi$ of the other EIT field with respect to the atomic resonance. The power of the beams was   115 $\mu$W, with spot sizes of $w\simeq 1.7 \,mm$ ($I\simeq0.2\,I_{sat}$). The repump beam  was kept on resonance with a  power of $\sim$ 750 $\mu$W. 

The transmitted EIT beams are sent to amplified photodetectors, with a signal proportional to the  mean intensity (DC) and a high gain transimpedance amplifier  for high frequency components (HF). The HF signals are demodulated by an electronic oscillator at 2 MHz, and the low-pass filtered outputs ($f_c<300$ kHz) are acquired by an analog-to-digital converter, together with the DC signals.

The measurement cycle is shown in Fig.\ref{fig:setup}(Bottom). After a MOT loading cycle of 900 ms,  the current of the trapping coils was turned off 3 ms before the measurement to avoid spurious magnetic fields that could lead to decoherence or energy shifts between the Zeeman ground-state levels. 
The cooling beams, exciting the D2 line, could be either turned on or off during the experiment. In order to pump the atoms to the $5S_{1/2}(F=1)$ level, the repump field of the MOT was turned off and the EIT repump turned on at the beginning of the acquisition.
During the measurements, the frequency detuning of one of the spectroscopy beams ($\delta'_2$) could be left at a fixed value, or could be scanned in a 15 MHz range (2.5 times the atomic linewidth of 6 MHz) centered in the two-photon resonance, during 0.5 ms which correspond to ${d\delta'/dt}=30$ MHz/ms. 
There are two limits for this scanning rate. The upper limit is given by the atomic evolution of the states, of the order of 30 $\mu$s, evaluated by dynamical Bloch equations. The scan rate should be low enough to satisfy an adiabatic evolution. On the other hand,  the atomic loss due to ballistic expansion gives a lower limit for the scanning time. In the current case, the 10\% loss for 1 ms can be considered as a reasonable value.
All the presented results are an average of 100 runs of the experiment.

%This condition is satisfied in our experiment, since the evolution of the atomic populations has a typical  time of 30 $\mu$s, evaluated by dynamical Bloch equations,  being at least one order of magnitude smaller than our acquisition time.

%\section{Experimental Results}

\begin{figure}[h!]
\hspace{-0.6cm}
\includegraphics[width=80mm]{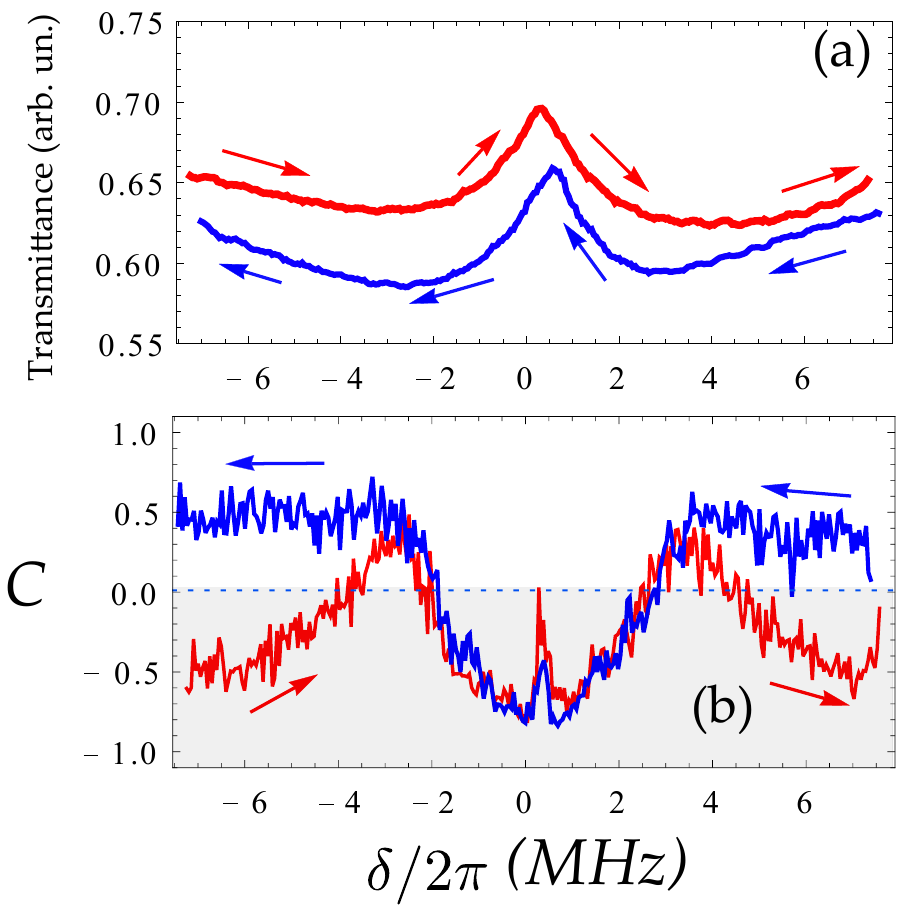}
\caption{(color online) Intensity (a)  and correlation spectra (b) for increasing (red) and decreasing (blue) scanning in frequency. The arrows indicate the time evolution of the scanning.}
\label{fig:Assy}
\end{figure}

Typical results for the noise correlation spectroscopy are shown in Fig.  \ref{fig:Assy}, for a fixed frequency detuning of field 1 ($\delta'_1=(0.0\pm0.5)$ MHz), comparing the transmitted intensity of the probe laser and the correlation evaluated according to Eq.~\ref{eq:Ceq}.
There is a relevant difference in the behavior of the correlation (Fig. \ref{fig:Assy}b) depending on the sign of the scanning rate $d\delta'/dt$. For a positive scan in frequency (red line), the curve shows anti-correlation out of the EIT region, while for negative scan (blue line) there is correlation. On the other hand, no significant difference is  observed for the mean transmitted  power (Fig. \ref{fig:Assy}a).
This asymmetry associated to the temporal evolution of the measurement is unexpected, since the sign of the scanning rate ${d\delta'/dt}$ should not be relevant  
for an adiabatic scan.

%Given those observations, we perform a step scan to avoid the scanning effect into the correlation spectra. This was done by fixing the value of detuning $\delta$  for each reloading cycle of the MOT,  recording the correlation spectrum for $0.5$ ms. Each of this measurements points is averaged over 200 spectrum samples.
%Figure \ref{fig:Assy}(c) shows the intensity correlation for the step scan.
% The spectrum obtained follows the increasing scan curve for $\delta < 0$  (Red detuned values) and the decreasing scan curve for $\delta > 0$ (Blue detuned values).  From this spectrum, we observe that the system is sensitive for the detuning region even in non-scanning conditions and therefore we affirm that the effects observed could not be attributed to scanning speed, unlike the results in Ref. ~\cite{ Greenberg09}. 
 %We also addressed the question of atomic loss related to the ballistic expansion of the cloud. Fluorescence and time of flight (TOF) measurements indicated that the population lost in the window of measurement were not significant. 

It should be noticed that the resulting spectrum is characterized by two particular regions. In the first one,  close to the EIT resonance ($|\delta'|< 3$ MHz), atoms are pumped and trapped in the dark-state and no asymmetry is observed, in accordance with our previous observations ~\cite{Florez16_p1}.
 The situation is different beyond the transparency window ($|\delta'| > 3$ MHz), where the mismatch in correlation spectra for increasing and decreasing scanning is very distinctive.
Therefore, the asymmetry is observed whenever the atoms are not in the dark state and can be excited by the laser beams.

\begin{figure}[h!]
\includegraphics[width=80mm]{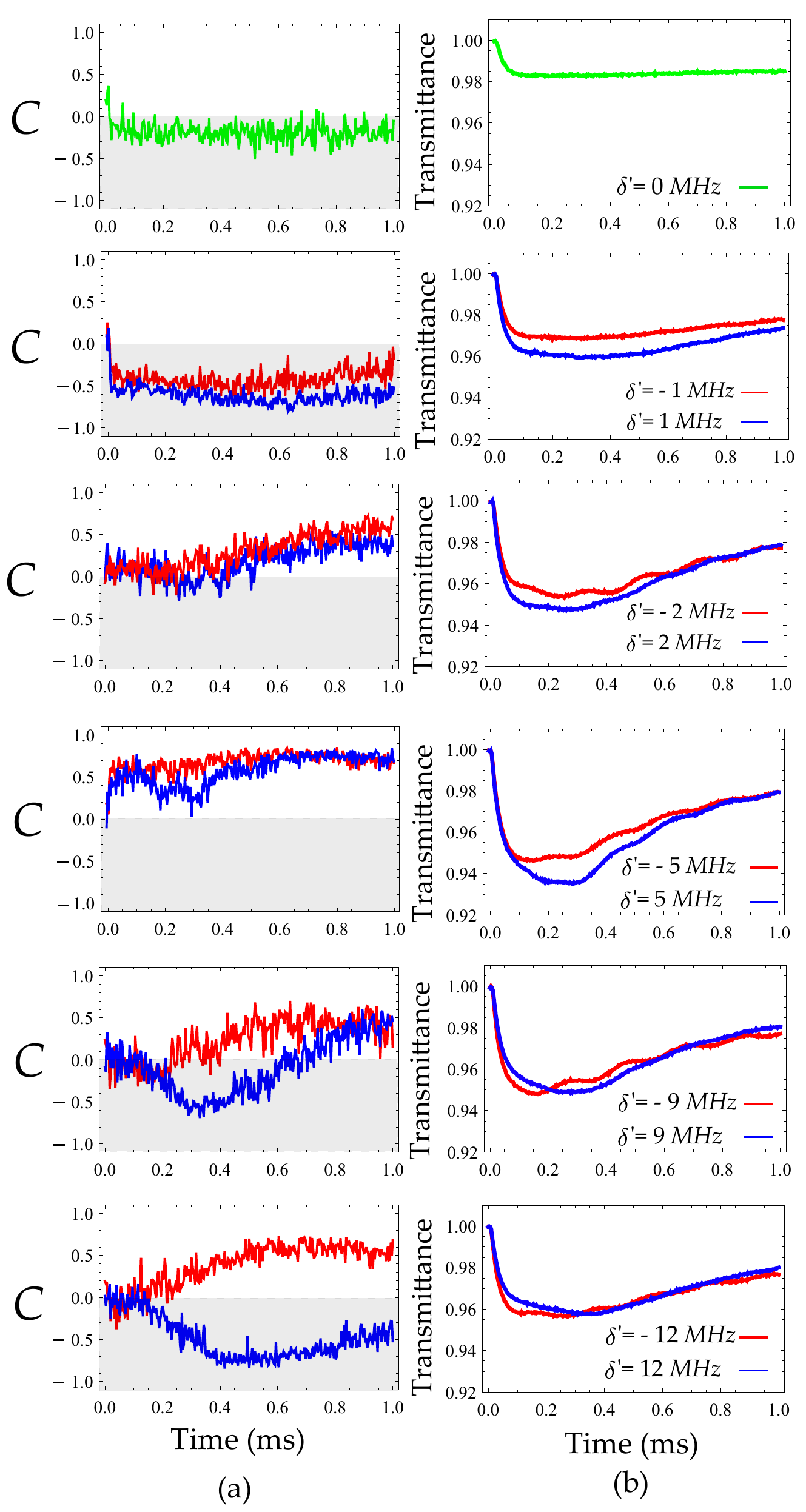}
\caption{(color online) Measurements of the time evolution of correlation function (first column) and transmitted power (second column) for different detuning frequencies $\delta'=\delta/2\pi$.}
\label{fig:TimeCorre}
\end{figure}

In order to  understand this asymmetry, we investigated the evolution of the system for different values of laser detuning frequency $\delta'$. For a fixed frequency of the probe,  the  correlation and transmission of the fields are measured for 1 ms, while the cooling beams and magnetic fields are turned off. The time scale of this measurement was found to be sufficient to observe the evolution of these measurements while the change of optical density due to ballistic expansion was smaller than 10\%. The results are presented  in Fig.~\ref{fig:TimeCorre}.

For the exact EIT condition ($\delta'= 0 $ MHz) the results of intensity and noise reach a steady value after about 40 $\mu$s,  consistent with the fast evolution of the atomic populations. Beyond that point, transmitted intensity has a very slow evolution, with a small increase in the transmittance ($< 0.5$ \%), consistent with changes in the atomic density.

As the detuning grows, the transmittance  presents a fast reduction, associated to the absorption of the field. For $\delta'=\pm1$ MHz this drop is still small, consistent with the fact  that the beam frequency is within the EIT resonance. The correlations show a fast drop to negative values, consistent with the observed values in Fig. \ref{fig:Assy}. Any further evolution is slow under the measured time. Nevertheless, a small asymmetry in  transmittance is observed. Similar results are observed for  $\delta'=\pm2$ MHz, with the correlation evolving to positive values.

Out of the EIT peak, but within the atomic linewidth, for $\delta'=\pm5$ MHz, absorption reaches its peak value. It is noticeable that the correlation signal presents a drop for positive detuning, at 0.3 ms. This diference in the evolution of the correlation signal according to the detuning becomes increasingly relevant as the detuning grows ($\delta'=\pm 9$ MHz), changing back and forth to negative values.
For $\delta'=\pm 12$ MHz, the discrepancy in the correlation signal is large enough to prevent the anticorrelation from recovering to positive values within the measurement time.

The slow temporal dependence is, therefore, present only when the atoms are out of the exact two-photon resonance. It is  consistent with the idea that light scattering from the atoms plays a role in the evolution of the signal. We propose a model where the resulting force from the absorption accelerates the atomic ensemble, leading to a time dependent Doppler shift of the atomic transitions that can be verified in the evolution of the measured fields.
%\section{Theoretical Model}
We  describe the atom-light interaction in terms of the two dynamical processes in our system, one  regarding the external degrees of freedom (atomic velocities) and the other one involving  the internal degrees of freedom (atomic states).
For the internal degrees of freedom along with the noise correlation we follow the calculation presented in ref.  \cite{Florez16_p1}. 
In what follows we briefly review the main aspects. 

Consider a three level system interacting with two
light fields $\mathbf{E}_1$ and $\mathbf{E}_2$ with phase noise in  a $\Lambda$-EIT  configuration. 
The phases of each field ($\phi_1(t)$ and $\phi_2(t)$) present random fluctuations,
following the statistics of a Wiener process, i.e. $\left\langle d\phi_i(t)\right\rangle=0$ and $
\left\langle d\phi_i(t) d\phi_j(t) \right\rangle = 2\gamma dt$ with $i,j=\{1,2\}$,   where $\left\langle \cdots \right\rangle$ stands for the statistical average and $\gamma$ corresponds to the spectral linewidth of the laser. One important point in the current experiment comes from the fact that both fields come from  the same laser, therefore $ \phi_1(t)=\phi_2(t)=\phi(t)$.

\begin{figure}[thb!]
\includegraphics[width=80mm]{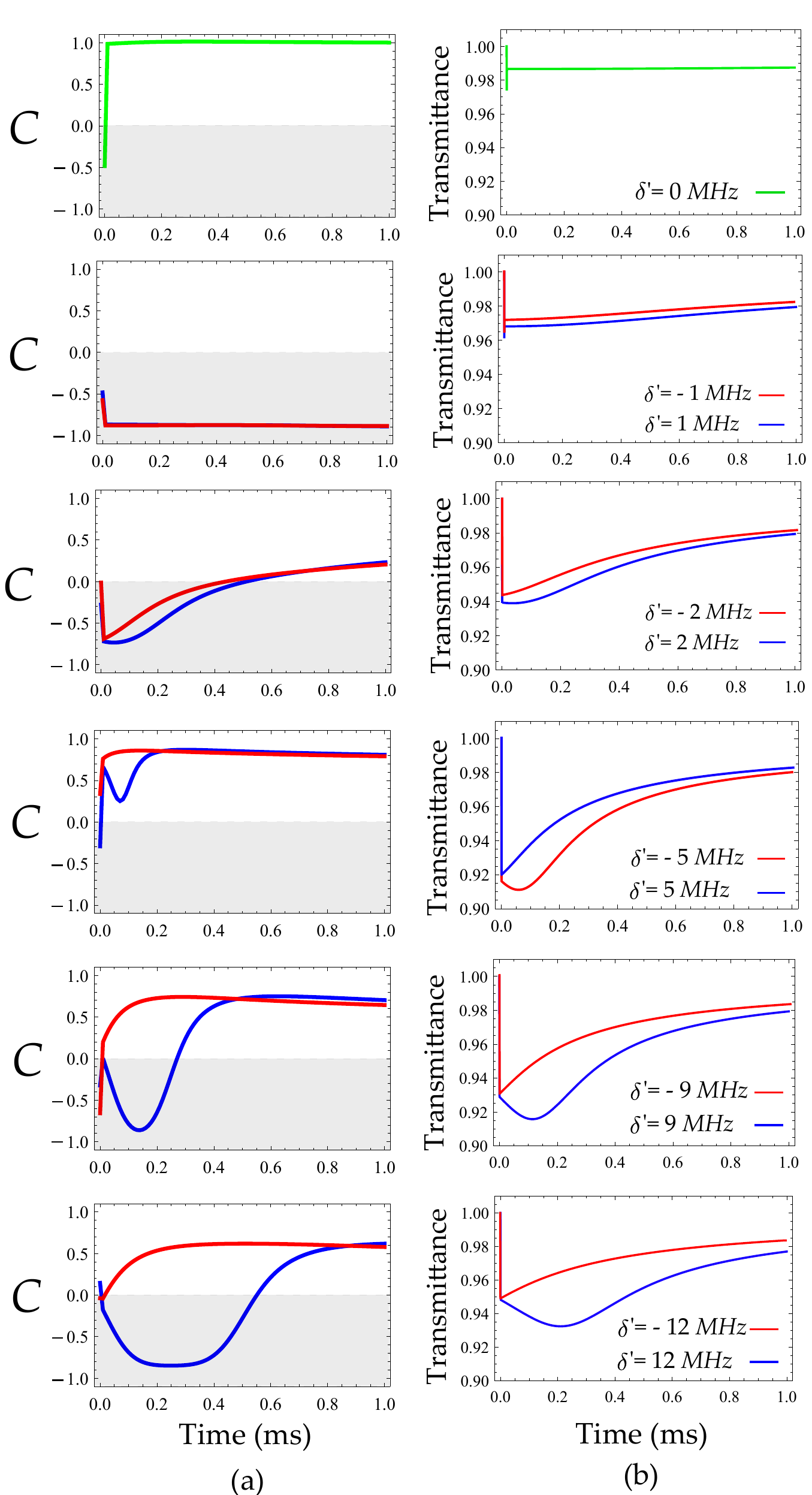}
\caption{(color online) Calculations for the time evolution of correlation function (first column) and transmitted power (second column) for different detuning frequencies $\delta'=\delta/2\pi$.}
\label{fig:Theoteo}
\end{figure}

The dynamics of the two fields coupling the three level system is given by 
the  Bloch equations in which the atomic density matrix is defined by the slow varying vector
$\mathbf{x}=(\rho_{11},\rho_{22},\rho_{13},\rho_{31},\rho_{23},\rho_{32},\rho_{12},\rho_{21})$ 
describing the atomic state.
According to  \cite{Florez16_p1} the phase sensitive Bloch equation is
\begin{eqnarray}
d\mathbf{x}(t)=\mathbf{M}\left[dt,d\phi^2,\mathbf{v}\right]\mathbf{x}(t)+\mathbf{B}\left[\mathbf{x}\right]d\phi(t)+\mathbf{x}_0 dt  \label{x_ave}.
\end{eqnarray}
Here vector $\mathbf{B}$, proportional to $\mathbf{x}$, does not contribute to the calculations given that $\left\langle d\phi \right\rangle=0$. The matrix $\mathbf{M}$ depends linearly on $dt,d\phi^2$ and contains the Rabi frequencies, the detunings and the spontaneous emission rate. In the present case, it also takes into account the contribution of the Doppler shift on the field detuning,  given by  $\mathbf{k}\cdot\mathbf{v}(t)=\mathbf{k}\cdot d\mathbf{r}(t)/dt$. 
Finally, the constant  vector $\mathbf{x}_0$ is  associated to the conservation of the atomic population.

In order to describe the dynamical equation for the velocity of the atoms, we assumed the dipole and long wavelength approximations, i.e. $\lambda \gg a$ (where $\lambda$ is the optical wavelength and $a$ the Bohr radius).
Each field exerts scattering forces. For a traveling wave, under the rotating wave approximation \cite{ref:metcalf},  $\bar{\mathbf{F}}_i=2\hbar \mathbf{k}\  \Omega_{i}(\mathbf{r}_0)\text{Im}\ \tilde{\rho}_{i3}$
where $\mathbf{r}_0$ represents the position of the atomic center of mass and $\text{Im}\ \tilde{\rho}_{i3}$ gives the absorption of each beam.
Under the effect of those forces, the atoms are accelerated, gaining a Doppler shift 
 affecting the evolution of the internal atomic variables through Eq.(\ref{x_ave}).

We can determine the time evolution of the mean value of the atomic states $\left\langle \mathbf{x}(t)\right\rangle $ under the effect of the optical forces by numerically solving the coupled averaged equations of atomic motion (thus evaluating the atomic velocity) and Eq. (\ref{x_ave}). The values of $\left\langle \mathbf{x}(t)\right\rangle $  give the absorption and the dispersion of the fields associated to $\text{Im}\rho_{i3}$ and $\text{Re}\rho_{i3}$, respectively, for $i=1,2$.  The transmission for each beam is given by $T_i(t)=\exp(-\alpha\text{Im} [\mathbf{x}(t)]_{2i+1})$ for $i=1,2$ where $\alpha$ corresponds to the optical depth.

We use the stochastic treatment in \cite{Florez13} to calculate the evolution of the density matrix elements in Eq.(\ref{x_ave}). We determine 
the spectral densities $S_{ij}(\omega)$ in Eq.(\ref{eq:Noiseeq}) and then the correlation given by Eq. (\ref{eq:Ceq}).
It is worth noting that the approach in \cite{Florez13} for calculating the correlation in Eq.(\ref{eq:Ceq})
considers a steady state condition of the atomic states. 
Therefore, the evaluation of  $C(\omega)$ for each $\mathbf{x}(t)$ in Eq.(\ref{x_ave})
is only valid for an adiabatic situation, where  the evolution of the internal degrees of freedom is fast compared to the changes in atomic velocity. In this case the atomic state given by $\mathbf{x}(t)$ follows the slowly-varying steady state value of $\mathbf{M}\left[dt,d\phi^2,\mathbf{v}\right]$. 

Resulting calculations for the evolution of correlation and transmittance are shown in Fig. \ref{fig:Theoteo} for different detunings.
The differences in the time evolution of these signals for the blue and red detunings are evident. 
On the other hand, the resonant case of EIT presents no slow transient evolution, as expected. The correlation value  at exact resonance is positive and nearly unitary. %The evaluated results are in good agreement with  our experimental observations (Fig. \ref{fig:TimeCorre}), although the correlation at exact resonance is limited by the narrow linewidth of the central peak. Any small detuning in the experiment  leads to a degradation of the correlation. 
%\textit{Is worth mentioning that the transmittance calculated in Fig~\ref{fig:Theoteo} shows at least 2\% difference for opposite detuning, which in principle would allow to identify the different the effect of light in opposite detuning of the atomic resonance. However, Figure~\ref{fig:TimeCorre} shows that this 2\% between opposite detuning is barely noticeable, due to the noise accumulate by the fluctuation in atomic density for each run of the experiment.}

Out of the EIT condition ($|\delta'|>$ 2 MHz),  we observe that, for a red detuned probe field, the accelerating atoms lead to a growing detuning, and therefore to an increase in the transmittance. On the other hand, if the probe field is initially blue detuned, absorption increases as the atoms are brought into resonance, and transmittance is increased after they pass through this condition.
As for the evolution of the noise correlation signal, it can be understood by the phase-to-amplitude noise conversion observed with diode lasers \cite{Yabusaki91}.
Close to the atomic resonance, phase fluctuations are converted into intensity fluctuations by the atomic resonance, but the sign of the conversion term depends on the laser  detuning.
In the present two-fields spectroscopy, for the red detuned probe, the resulting push increases the absolute value of detuning for both fields, which become red detuned as well. Since these fields are derived from the same laser, noise conversion results in correlated intensities \cite{Florez16_p1}.
On the other hand, if field 2 is blue detuned, the increasing velocity of the atoms makes field 1, initially resonant, red detuned. As a consequence, the intensity correlations  become anti-correlated in a first moment. This situation does not last, because field 2 eventually becomes red detuned as well, due to the acceleration of the atoms, and their intensities become correlated at the end, as observed experimentally. 

A direct comparison of the experimental results (Fig.~\ref{fig:TimeCorre}) with the calculations presented in Fig.~\ref{fig:Theoteo} shows a good agreement for the correlation signal, and a limited agreement for the intensity measurements.
The measured transmittance is affected by the statistical noise introduced by fluctuations in atomic density, and the small change of transmittance for opposing detunings, of the order of  2\% (Fig~\ref{fig:Theoteo})
 is within the intensity fluctuations observed for different runs.
On the other hand, the wide change in the correlation makes this signal more robust against perturbations, leading to a better agreement of the observed values and the presented model, and an enhanced sensitivity to the effect of optical forces.
%The most noticeable discrepancy for correlation at exact resonance can be accounted for the narrow linewidth of the central peak. Any fluctuation of the detuning in the experiment  leads to a degradation of the correlation in the averaged value.
 
%Therefore,  since the correlation is insensitive to density fluctuations, the theoretical model is in good agreement with the experimental observation, showing the robust difference of the correlation for opposite detunings. 
%Therefore, the noise correlation detection enhance the sensitivity of light forces with respect to the standard transmittance detection.

With the knowledge of the evolution of the system under the effect of the radiation pressure over the atom out of the EIT region, we can understand the asymmetry observed in Fig. \ref{fig:Assy}. For the scanning with an increasing frequency (red line), the system may begin with an anti-correlated signal.
 As the lasers push the atoms, leading to a red shift for field 1, the probe laser (field 2) is scanned, ending with a blue detuning, and therefore the resulting situation leads back to anti-correlation for the fields. 
 On the other hand, for the scan with a negative variation of the probe frequency, we may begin with correlations for a blue detuned probe. As the atoms are pushed by both fields, at the end of the scan, both fields are red detuned and correlation shows up again after crossing the EIT region.
From the calculations in Fig.\ref{fig:Theoteo} we estimate that the acceleration due to light forces
%with which the atoms are pushed by the light 
is of the order of $10^3$ m/s$^2$ resulting in Doppler shifts of the order of the atomic line ($\simeq6$ MHz), or atomic velocities of 1 m/s (on the order of magnitude of the thermal velocity of our cold atom cloud).

%Therefore, care should be taken on the treatment of the noise correlation spectroscopy, regarding the information of the asymmetry involving frequency detunings external to the EIT central peak. As observed in \cite{Florez16_p1}, one-photon detuning of the probe peak shifts the correlation/anti-correlation relation at the sides of the peak. Nevertheless, the Doppler shift can induce such difference, mimicking a small detuning of the fields. A fast average of the measurements under short  time evolution can correct this effect, as observed in Fig.\ref{fig:Assy}c.

%\section{discussion}

\begin{figure}[h!]
\includegraphics[width=70mm]{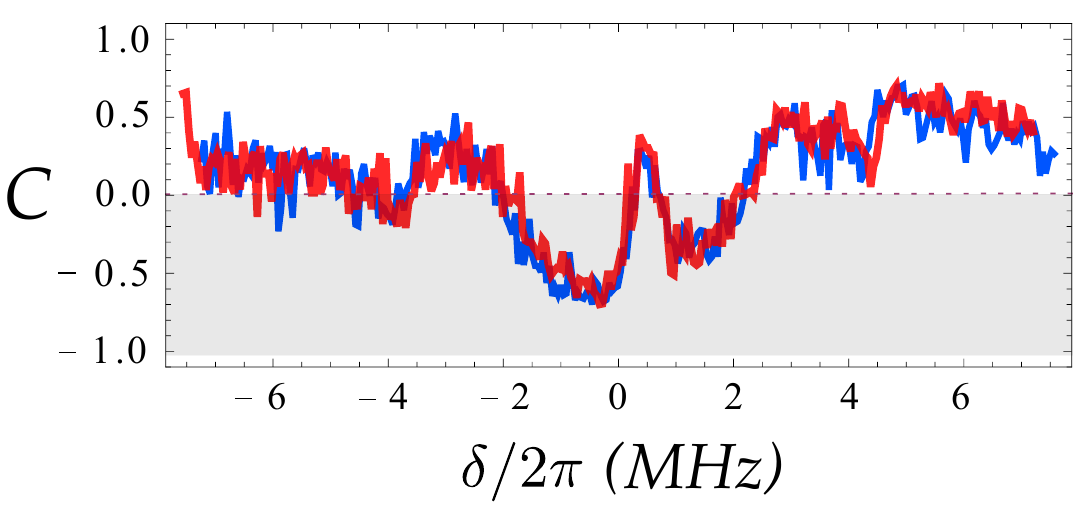}
\caption{(color online) Correlation spectra for increasing (red)
and decreasing (blue) scanning in frequency for the counter propagating configuration of the EIT beams.}
\label{fig:counterProp}
\end{figure}

One way to reduce the effect of the radiative pressure is to perform a spectroscopy in a counter-propagating configuration for the EIT beams. Therefore, the forces are nearly balanced ($\bar{\mathbf{F}}_1-\bar{\mathbf{F}}_2\sim0$) and atomic acceleration is negligible. As a consequence, the correlation spectra (Fig. \ref{fig:counterProp}) present no asymmetry with respect to the scanning direction. 

%\section{Conclusions}
In conclusion, the developed model gives a good agreement with the observed results, successfully coupling internal and external degrees of freedom of the atoms.
We can observe that the sensitivity of the noise spectroscopy allows the observation of the pushing of the atoms in the vicinity of the EIT condition. 
Although the intensity spectrum presents a distinctive feature at different detunings, as observed in the theory (Fig.~\ref{fig:Theoteo}), its observation is limited by atomic density fluctuations, while this effect is much more evident  in the correlation spectroscopy, with a swing from anti-correlation to correlation of the relevant signal (Fig.~\ref{fig:TimeCorre}).  
Therefore, correlation can become a measurement tool for the evaluation of atomic acceleration under optical forces. 

% (in this particular case the transient effect is due to Doppler effect from light forces).
Moreover, under counter-propagating geometry, the contribution of the optical forces are compensated, thus  the noise correlation could be employed to detect additional external forces (electric, magnetic or grativational), as well as to evaluate their slow drift in time, which would translate into optical detuning shifts.

The authors acknowledge  support from grant \# 2010/08448-2, \href{http://dx.doi.org/10.13039/501100001807}{Funda\c c\~ao de Amparo \`a Pesquisa do Estado de S\~ao Paulo (FAPESP)},

%\FloatBarrier

\end{document}